# CONFIRMATION OF THE OGLE-2005-BLG-169 PLANET SIGNATURE AND ITS CHARACTERISTICS WITH LENS–SOURCE PROPER MOTION DETECTION

V. Batista[1], J.-P. Beaulieu[1], D. P. Bennett[2], A. Gould[3], J.-B. Marquette[1], A. Fukui[4], and A. Bhattacharya[2]
[1] UPMC-CNRS, UMR 7095, Institut d'Astrophysique de Paris, 98Bis Boulevard Arago, F-75014 Paris, France; batista@iap.fr
[2] University of Notre Dame, Department of Physics, 225 Nieuwland Science Hall, Notre Dame, IN 46556-5670, USA
[3] Department of Astronomy, Ohio State University, 140 West 18th Avenue, Columbus, OH 43210, USA
[4] Okayama Astrophysical Observatory, National Astronomical Observatory of Japan, Asakuchi, Okayama 719-0232, Japan

## ABSTRACT

We present Keck NIRC2 high angular resolution adaptive optics observations of the microlensing event OGLE-2005-BLG-169Lb, taken 8.21 years after the discovery of this planetary system. For the first time for a microlensing planetary event, the source and the lens are completely resolved, providing a precise measurement of their heliocentric relative proper motion, $\mu_{\rm rel,helio} = 7.44 \pm 0.17$ mas yr$^{-1}$. This confirms and refines the initial model presented in the discovery paper and rules out a range of solutions that were allowed by the microlensing light curve. This is also the first time that parameters derived from a microlensing planetary signal are confirmed, both with the Keck measurements, presented in this paper, and independent measurements obtained with the *Hubble Space Telescope* in *I*, *V* and *B* bands, presented in a companion paper. Hence, this new measurement of $\mu_{\rm rel,helio}$, as well as the measured brightness of the lens in *H* band, enabled the mass and distance of the system to be updated: a Uranus-mass planet ($m_{\rm p} = 13.2 \pm 1.3 M_\oplus$) orbiting a K5-type main sequence star ($M_* = 0.65 \pm 0.05 M_\odot$) separated by $a_\perp = 3.4 \pm 0.3$ AU, at the distance $D_{\rm L} = 4.0 \pm 0.4$ kpc from us.

## 1. INTRODUCTION

Gravitational microlensing probes exoplanets down to a few Earth masses (Bennett & Rhie 1996) beyond the snow line, where core accretion theory predicts that most planets will form (Lissauer 1993; Ida & Lin 2004). The method does not depend on the luminosity of the host star. Therefore, the sensitivity extends to planets orbiting any kind of star in the galactic disk to the galactic bulge (Janczak et al. 2010). When the source and the lens stars are aligned to better than ≲1 mas, the flux of the source star is magnified. A companion to the lens star might reveal its presence by additional perturbations to the observed photometric light curve (Mao & Paczyński 1991; Gould & Loeb 1992; Griest & Safizadeh 1998).

Microlensing parameters of the lens star and its companion(s) can be securely deduced from the modeling, such as their mass ratio and separation in units of Einstein radius $\theta_{\rm E}$, but the physical parameters of the system often remain uncertain due to a lack of information about the lens, unless some second-order effects are measured, such as parallax and finite-source effects (Gaudi et al. 2008; Dong et al. 2009; Batista et al. 2011; Muraki et al. 2011; Han et al. 2013). Otherwise, there is only a single measurable parameter, the Einstein radius crossing time, $t_{\rm E}$, to constrain the lens mass, distance, and the relative lens–source proper motion, $\mu_{\rm rel}$. In such a case, one usually makes a Bayesian analysis based on Galactic models to estimate the physical properties of the system (Beaulieu et al. 2006; Batista et al. 2011). A complementary method to constrain the lens mass consists in measuring its flux directly (or an upper limit) using high angular resolution imaging. High angular resolution allows us to resolve the source star from their unrelated neighbors, although the images of the source and lens stars will generally remain blended together for years. The microlensing models determine the *H*-band brightness of the source star from *H*-band light curves, so it is possible to determine the *H*-band brightness of the host star (lens) by subtracting the source flux from the *H*-band measurement of the combined host+source flux. This flux can be obtained with a large ground-based telescope using adaptive optics (AO) such as Keck (Batista et al. 2014 and references therein).

A complete understanding of the physical properties of the microlensing systems is fully possible when the source and the lens are separated enough (~50–60 mas) on the Keck images. Since the relative proper motion is typically in the range 4–8 mas yr$^{-1}$, we would have to wait 5–10 years after the microlensing event to perform such measurement. Indeed this enables both the measurement of the lens and the source fluxes independently of any modeling. It also provides a measurement of the amplitude and direction of the lens–source relative proper motion. In most cases, it will break the potential remaining degeneracies in the modeling and will provide the lens mass and distance unambiguously. For the important subclass of events with the so-called "one-dimensional microlens parallax" measurement, high angular resolution can provide even tighter cross checks (Gould 2014).

One of the best candidates to perform such a measurement with is the system with a Uranus-mass planet detected in the microlensing event OGLE-2005-BLG-169 (Gould et al. 2006). The fitted relative lens–source proper motion, and the predicted lens/source flux ratio makes it an ideal candidate (Bennett et al. 2007). Indeed, the lens was expected to be easily detected few years after the event with high angular resolution imaging, especially since the source is a main sequence star (compared to a scenario where the source would be a giant star) and the lens is likely to be brighter than 16% of the combined lens +source flux. We took high angular resolution images with the Keck II telescope in 2013 July, i.e., more than eight years after the microlensing event, and we measured the lens–source relative proper motion, $\mu_{\rm rel,helio}$, since the two objects were well separated at that time. We also obtained a precise measurement of the flux ratio between the lens and the source, $f_{\rm L}/f_{\rm S}$, in $H$ band. This is the first time that a proper motion determined from the planetary signal of a microlensing light curve is confirmed by direct measurement, with a companion paper, Bennett et al. (2015), presenting *Hubble Space Telescope* (*HST*) observations of the same system (in $I$, $V$, and $B$ bands).

We perform a Bayesian analysis using the constraints from both the 2005 microlensing light curve ($t_E$, $\theta_E$, $f_S$) and the Keck measurements ($\mu_{\rm rel,helio}$, $f_L/f_S$) as prior distributions, as well as a mass–luminosity function for main-sequence stars (isochrones from An et al. 2007 and Girardi et al. 2002) to determine the properties of the lensing system with the highest probability. We will compare our values to the predictions from the light curve modeling obtained in 2005 (Gould et al. 2006; Bennett et al. 2015). This Keck measurement of the lens–source relative proper motion enables the error bar on the lens mass to be reduced by a factor of 6. It yields a complete solution of the microlensing event and allows us to determine the system properties in physical units (distances and masses). This follow-up technique with high angular resolution will be applied to other microlensing events in which the host star mass is not well constrained, and whose lens is likely to be detected.

## 2. OGLE-2005-BLG-169 LIGHT CURVE DATA AND MODELS

OGLE-2005-BLG-169Lb is a planetary system composed of a Uranus-like planet orbiting a $\sim 0.5\, M_\odot$ star, discovered in 2005 and published by Gould et al. (2006). It manifested during a high magnification microlensing event ($A \sim 800$) that was alerted by the OGLE collaboration in 2005 April and followed by the $\mu$FUN, RoboNet, and Planet collaborations. One of the telescopes that collected data on this event was the 1.3 m SMARTS telescope at CTIO, which is equipped with the ANDICAM, taking images simultaneously in the optical and infrared. The $H$-band data were not used in the original analysis (Gould et al. 2006), but they are important here because they allow us to determine the $H$-band brightness of the source star. The $H$-band light curve also allows us to determine the source radius more accurately because stellar color–radius relations are more precise since they use optical minus infrared colors (Kervella et al. 2004; Boyajian et al. 2014). Also, as discussed in Bennett et al. (2015), an inconsistency was discovered between the CTIO $I$-band and OGLE $I$-band photometry used in the original paper. As a result, data from all three of the CTIO passbands ($V$, $I$, and $H$) were reduced with SoDoPHOT (Bennett et al. 1993), replacing the original DoPHOT reductions used in Gould et al. (2006). These new reductions resolved the discrepancy between the OGLE and CTIO $I$-band reductions.

As Figure 1 (left panel) indicates, only the second half of the planetary feature is covered by high cadence observations. The observations are sparse on the rising side of the light curve. In contrast, the coverage on the falling side of the peak is very dense, with observations every 10 s from the 2.4 m MDM telescope for a period of 3 hr. It is these MDM data that contain the OGLE-2005-BLG-169Lb planetary signal, and it is only the high cadence and precision of the MDM data that allowed a planet to be detected with such a low amplitude ($\sim 2\%$) signal. The discovery paper (Gould et al. 2006) explored the implications of this incomplete light curve coverage, as well as the possibility of systematic errors in the MDM photometry. They found that the sparse early light curve coverage led to some ambiguity in the implied light curve parameters. The best fit was found to be at $(q, \alpha) = (8 \times 10^{-5}, 118°)$, where $q$ is the planet–star mass ratio and $\alpha$ is the angle of the lens trajectory with respect to the lens axis. However, there were several other local $\chi^2$ minima that were acceptable solutions: $(q, \alpha) = (6 \times 10^{-5}, 88°)$ (the second best solution) and $(q, \alpha) = (7 \times 10^{-5}, 103°)$. While these second and third best solutions exist for both the Stanek and OGLE pipeline reductions of the MDM data, they are disfavored by a smaller $\Delta\chi^2$ for the Stanek reductions.

These different solutions do not predict identical $\mu_{\rm rel,geo}$ values, due primarily to the variation in the angle $\alpha$. Since the source crosses the caustic curve where it is nearly parallel to the lens axis, the solutions with $\alpha \approx 90°$ have a source trajectory that crosses the caustic at an angle that is nearly perpendicular, while the solutions with $\alpha \approx 30°$ have a caustic-crossing angle of $\sim 60°$ (see Figure 2). Since the duration of the caustic crossing is measured by the MDM data, these models predict source radius crossing time ($t_*$) values that differ by a factor of $\sim \cos 60° = 0.866$. In particular, the second best models with $\alpha \simeq 88°$ have $t_*$ values that are larger than the $t_*$ values for the models with $\alpha \simeq 118°$ by about this factor of $1/0.866$. This translates into a geocentric lens–source relative proper motion $\mu_{\rm rel,geo} = \theta_*/t_*$ that is smaller by the same factor of $\sim 0.866$, so that $\mu_{\rm rel,geo} \approx 7.3$ mas yr$^{-1}$ would be preferred by the second best model over the $\mu_{\rm rel,geo} \approx 8.4$ mas yr$^{-1}$ value predicted by the best fit model. This higher value of $\mu_{\rm rel}$ is also the one presented in Gould et al. (2006), where they measured the radius of the source from the caustic-crossing features of the light curve, $\rho = 4.4^{+0.9}_{-0.6} \times 10^{-4}$, and $\theta_* = 0.44 \pm 0.04\, \mu$as, using the standard Yoo et al. (2004) approach, yielding $\theta_E = \theta_*/\rho = 1.00 \pm 0.22$ mas and $\mu_{\rm rel} = \theta_E/t_E = 8.4 \pm 1.7$ mas yr$^{-1}$ ($3\sigma$ ranges). Combined with a Bayesian analysis using a Galactic model and the weak parallax constraint as a prior, they derived a mass and distance for the lens: $M_{\rm host} = 0.40^{+0.23}_{-0.29}\, M_\odot$ and $D_L = 2.7^{+1.6}_{-1.3}$ kpc.

With the new CTIO photometry, and the Stanek reduction of the MDM photometry, this $\alpha \simeq 88°$ model that was previously 2nd best, now has a $\chi^2$ that is slightly better than the $\alpha \simeq 118°$ model. This best fit model with $\alpha \simeq 88°$ is shown in Figure 1 (right panel). As Gould et al. (2006) found, the light curve is consistent with both the $\alpha \simeq 88°$ and $\alpha \simeq 118°$ models. But, our Keck measurements of $\mu_{\rm rel}$ are not consistent with all these different $\alpha$ values. The measured $\mu_{\rm rel}$ value is at the lower edge of the range that is consistent with the light curve data. As a result, the models with $\alpha \simeq 118°$ and $\alpha \simeq 103°$ are now excluded because they require smaller $t_*$ values which imply $\mu_{\rm rel}$ values that are inconsistent with the Keck measurement of $\mu_{\rm rel}$. This Keck constraint on $\mu_{\rm rel}$ also tightens on the constraints on other parameters, such as the mass ratio $q$, which is now $q = 5.7 \pm 0.4 \times 10^{-5}$ instead of $q = 8^{+2}_{-3} \times 10^{-5}$.

The microlensing parameters that we consider as prior distributions in the Bayesian analysis presented in this paper are given in Table 1. The two microlensing models described previously are labeled as Gould$_{2006}$ and Bennett$_{2015}$. We will use the parameters from the updated model Bennett$_{2015}$ and compare our final results with the conclusions of Gould et al. (2006).

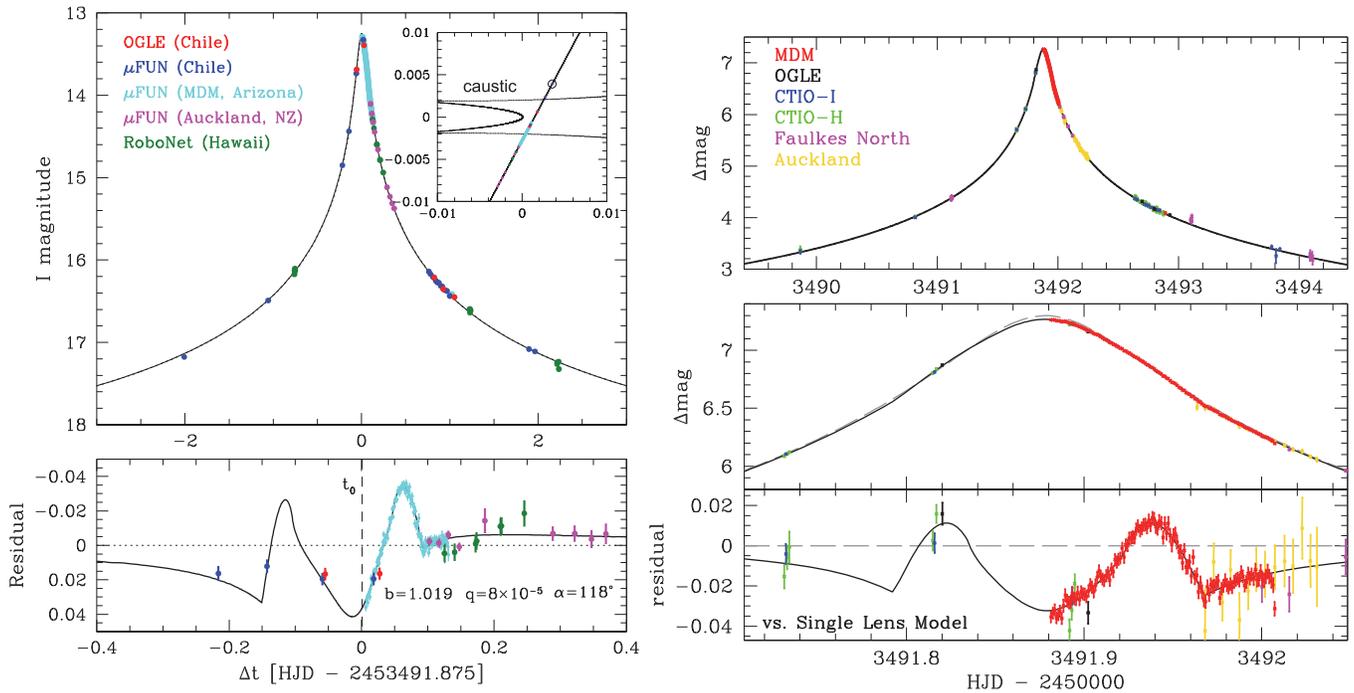

**Figure 1.** Left: light curve of OGLE-2005-BLG-169 from Gould et al. (2006). Top: data and best-fit model. Bottom: difference between this model and a single-lens model with the same ($t_0$, $u_0$, $t_E$, $\rho$). Data from different observatories are represented by different colors; see the legend. Right: the light curve peak of event OGLE-2005-BLG-169 with the new CTIO photometry and the Stanek reductions of the MDM data. The best-fit model is indicated by the black curve, and the gray dashed curve indicates the same model without the planetary signal. The bottom panel shows the residual with respect to this no-planet model. This model is consistent with the Keck $\mu_{\rm rel,helio}$ measurement, while the model presented in the left panel model is not.

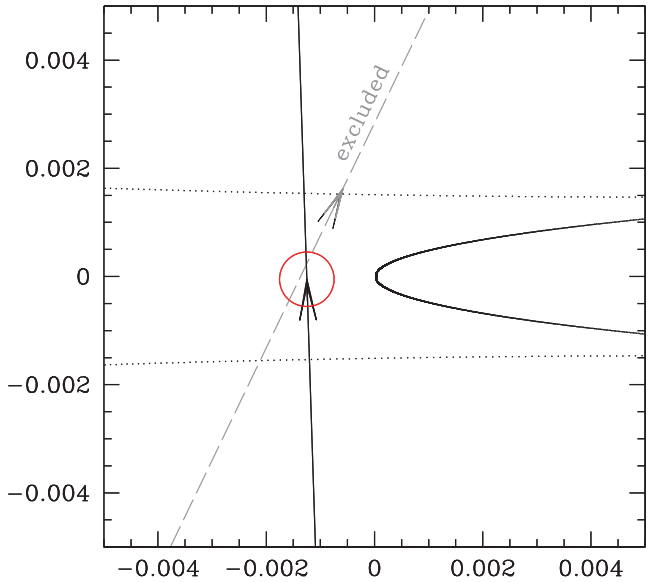

**Figure 2.** Caustic configuration for the OGLE-2005-BLG-169 model shown in Figure 1. The black line, with arrow, shows the source trajectory for this model, while the gray dashed line shows the source trajectory for the other local $\chi^2$ minimum for the light curve modeling. These models are consistent with the light curve, but they are contradicted by the relative proper motion measurement.

### 3. GOING FROM STAR/PLANET MASS RATIOS TO PHYSICAL MASSES

As explained in Batista et al. (2014), for most microlensing events, there is only a single measurable parameter, the Einstein radius crossing time, $t_E$, to constrain the lens mass,

**Table 1**
Model Parameters from the Microlensing Light Curve

| Parameter | Units | Bennett$_{2015}$ | Gould$_{2006}$ |
|---|---|---|---|
| $t_E$ | days | $41.8 \pm 2.9$ | $43 \pm 4$ |
| $\theta_E$ | mas | $0.965 \pm 0.094$ | $1.00 \pm 0.22$ |
| $H_S$ | ... | $18.81 \pm 0.08$ | $18.83 \pm 0.09$ |
| $t_*$ | days | $0.0202 \pm 0.0017$ | $0.019 \pm 0.004$ |

distance, and the relative lens–source proper motion, $\mu_{\rm rel}$, where:

$$t_E = \frac{\theta_E}{\mu_{\rm rel,geo}}, \quad \theta_E^2 = \kappa M_L \pi_{\rm rel},$$

$$\pi_{\rm rel} = {\rm AU}\left(\frac{1}{D_L} - \frac{1}{D_S}\right) \text{ and}$$

$$\kappa = \frac{4G}{c^2 {\rm AU}} = 8.144 \text{ mas } M_\odot^{-1}. \quad (1)$$

In addition, it is often possible to detect finite source effects in the light curve and to measure the angular size of the source. This leads to the measurement of the source radius crossing time, $t_*$, which provides an estimate of $\mu_{\rm rel}$ if one can measure the radius of the source, $\theta_*$, from its brightness and color, $\mu_{\rm rel} = \theta_*/t_*$. Hence, we can usually derive a relation between the mass of the host star and its distance from observables:

$$M_L \pi_{\rm rel} = \frac{\theta_*^2 t_E^2}{\kappa t_*^2}. \quad (2)$$

The "microlens parallax," $\pi_E$, expressed by

$$\pi_E = \sqrt{\frac{\pi_{rel}}{\kappa M_L}}, \quad (3)$$

as well as $\theta_E$, when measured, enables $M_L$ and $\pi_{rel}$ to be derived:

$$M_L = \frac{\theta_E}{\kappa \pi_E} \quad \text{and} \quad \pi_{rel} = \theta_E \pi_E. \quad (4)$$

However, without $\pi_E$, we usually cannot do better than the relation in Equation (2), which leaves one variable unconstrained. Note that the source distance is usually quite well known, so that measuring $\pi_{rel}$ is virtually equivalent to determining $D_L$.

In most cases it is possible in principle to detect and study (or to put upper limits on) the host (lens) star using high angular resolution images, either from space or ground-based AO observations. Keck's high angular resolution allows us to resolve the source+lens stars from their unrelated neighbors, and in the case of OGLE-2005-BLG-169, it also enables the lens to be separated from the source since the images were taken 8.2122 years after the microlensing event.

When there are magnified $H$-band data taken during the microlensing event (as in the present case), the microlensing model determines the $H$-band brightness of the source star, so it is possible to determine the $H$-band brightness of the host star (lens) by measuring the lens/source flux ratio on the Keck $JHK$-band images. Formally, this can be combined with Equation (2) and a $JHK$-band mass–luminosity relation to yield a unique solution for the host star mass. This would yield the planetary mass and star–planet separation in physical units because the planet–star mass ratio, $q$, and the separation in Einstein radius units are already known from the microlensing light curve.

Moreover, taking high angular resolution images many years after the microlensing event provides additional information if the source and the lens are resolved at that time. Indeed, it allows us to measure the heliocentric relative proper motion between the lens and the source, knowing their projected separation and the elapsed time. Furthermore, the geocentric relative proper motion is involved in the microlensing modeling as it can be deduced from the source trajectory through the caustic features created by the lens system. It also links the Einstein radius $\theta_E$ and the timescale $t_E$ of the microlensing event, $\mu_{rel} = \theta_E/t_E$. Therefore, being able to measure this parameter via an independent method such as AO observations is a robust way to test the validity of the microlensing models.

## 4. ADAPTIVE OPTICS OBSERVATIONS OF OGLE-2005-BLG-169

OGLE-2005-BLG-169 was observed with the NIRC2 AO system on the Keck telescope in $H$ band with natural guide star on 2013 July 18 (HJD = 2456491.388), with the narrow camera giving a plate scale of 0.01 arcsec pixel$^{-1}$. We chose five dithering positions with a step of 2 arcsec, 30 s exposure time for each, and we obtained 15 good quality images. We corrected for dark current and flatfielding using standard techniques. We then performed astrometry on the 15 frames and used SWARP from the Astromatics package to stack them. The detailed procedure is presented in Batista et al. (2014).

The stacked Keck image is shown in Figure 3 and clearly reveals an offset between the source and the lens positions. Their coordinates are (R.A., decl.)$_{source}$ = ($18^h06^m05^s.373$, $-30°43'58.03''$), (R.A., decl.)$_{lens}$ = ($18^h06^m05^s.377$, $-30°43'57.99''$).

The flux ratio $f_L/f_S = 1.75 \pm 0.03$ in $H$ band between the lens and the source has been measured with Starfinder (Diolaiti et al. 2000) as described in Kubas et al. (2012). This measurement is very robust and independent of calibrations.

Several arguments make unambiguous the identification of the lens and the source between these two stars. The astrometry of the Keck images was done using the $I$-band OGLE IV catalog. Another astrometry calibration was done on CTIO images taken during the peak of the microlensing event, when the source was still magnified, to detect its precise position (since $A \sim 800$). When comparing Keck to CTIO, the source position on CTIO images, (R.A., decl.) = ($18^h06^m05^s.38$, $-30°43'58.00''$), is very close to the position of the lower star of the detected Keck couple, i.e., the fainter one. Thus, a likely scenario is that this latter one is the source, since the foreground lens is probably moving faster than the background source. Moreover, the flux ratio between the blending light and the source from the light curve modeling is comparable to the flux ratio between the upper and the lower stars. The Keck images reveal the stars that are unresolved in the CTIO images, which could be responsible for the blending light in addition to the lens, and their contribution is very minor. Indeed, the only possible candidate is the faint star on the right on the target in Figure 3, which is four times fainter than the lens. It is then extremely likely that the lens is the brighter of the two stars (upper left). Finally, Bennett et al. (2015) confirm this identification on their $HST$ images, using information from three different passbands.

We already know the source flux from the modeling of the microlensing light curve with the $H$-band CTIO data. The updated model with the SoDoPhot data reduction (Bennett et al. 2015, Table 1) gives a source magnitude in $H$ band of $H_S = 18.81 \pm 0.08$. The flux ratio then implies a lens magnitude of $H_L = 18.20 \pm 0.10$.

This value is also consistent with the one we would have calculated from the Gould et al. (2006) model. The source magnitude in $I$ band is given from the light curve modeling with CTIO data, $I_{S,model} = 20.81 \pm 0.08$. Morever, the CTIO data in $H$ and $I$ bands lead to an instrumental source color of $(H - I)_{CTIO} = 2.024 \pm 0.011$ calculated by linear regression (i.e., without reference to models). Additionally, the calibration of CTIO by the 2MASS catalog gives an $I$-band zero point $\Delta I_{CTIO/2MASS} = 4.00 \pm 0.03$. This corresponds in the calibrated 2MASS system to $H_{S,model} = 18.83 \pm 0.09$ and then $H_L = 18.23 \pm 0.10$, which is very close to the value we use here.

Although the Keck field of view with the narrow camera is not large enough to perform a robust calibration due to the lack of comparison stars with 2MASS or even with data from the VVV survey (Minniti et al. 2010) done with the VISTA 4 m telescope at ESO, it is still interesting to estimate the lens flux with a method that is independent from the light curve modeling, as a sanity check. Hence, after having calibrated the VVV field with the 2MASS catalog, we use the two brightest stars in common with the Keck field to estimate the following magnitude for the source and the lens: $H_{S,2MASS} = 18.75 \pm 0.26$ and $H_{L,2MASS} = 18.15 \pm 0.26$. Having only two common stars for the calibration induces big error bars, but this measurement is in agreement with the previous one using only the flux ratio from the Keck image.

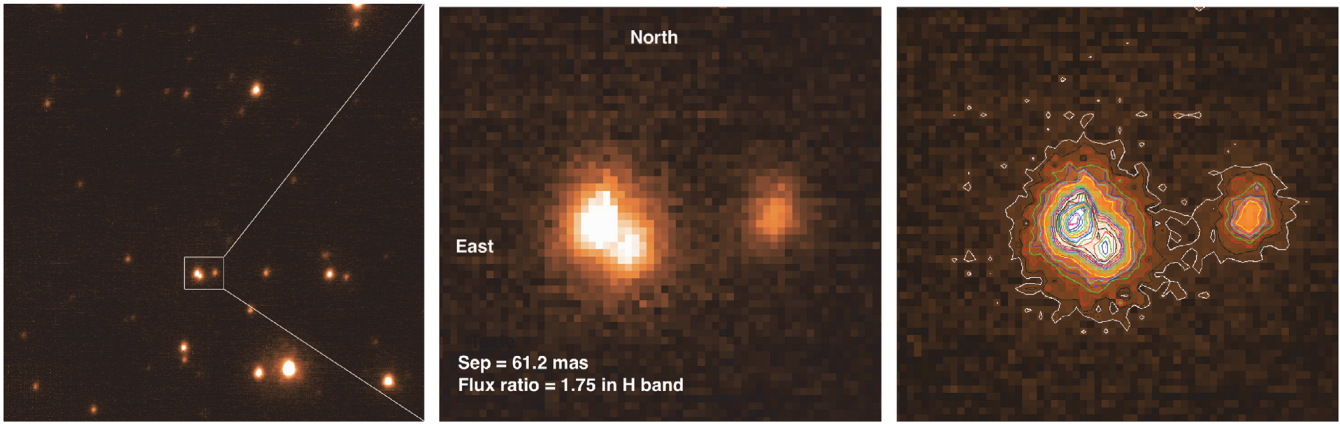

**Figure 3.** Left: Keck image of OGLE-2005-BLG-169 in $H$ band ($\sim 24'' \times 21''$). Middle: a zoom on the target showing the lens on the upper left and the source. They are separated by $\sim 61$ mas. The extra star on the right was part of the measured blending in the microlensing light curve. Right: zoom on the target showing the flux contours.

### 4.1. Lens–Source Relative Proper Motion

The lens and the source appear resolved on the Keck images (see Figure 3), 8.2122 years after the microlensing event was observed, at the following coordinates: (R.A., decl.)$_{\rm source}$ = ($18^{\rm h}06^{\rm m}05\overset{\rm s}{.}373, -30°43'58\overset{''}{.}03$),　(R.A., decl.)$_{\rm lens}$ = ($18^{\rm h}06^{\rm m}05\overset{\rm s}{.}377, -30°43'57\overset{''}{.}99$).

The offset between the two objects is ($\Delta$R.A., $\Delta$decl.) = ($0\overset{''}{.}046, 0\overset{''}{.}040$) $\pm$ ($0\overset{''}{.}001, 0\overset{''}{.}001$), i.e., a separation of $61.2 \pm 1.0$ mas. We multiplied by 4 the error bar given by Starfinder to be conservative. This would imply a heliocentric relative proper motion

$$\boldsymbol{\mu}_{\rm rel,helio}({\rm R.A.,\ decl.}) = (5.63, 4.87) \pm (0.12, 0.12) {\rm\ mas\ yr}^{-1},$$

i.e., $\mu_{\rm rel,helio} = 7.44 \pm 0.12$ mas yr$^{-1}$. We indeed consider the Keck measurement to be nearly expressed in the heliocentric frame because the difference is less than $\lesssim 0.2$% (i.e., $\lesssim \pi_{\rm rel}/\Delta\theta \sim 0.12$ mas/61 mas). In the Galactic coordinate system, we find

$$\boldsymbol{\mu}_{\rm rel,helio}(l, b) = (7.28, 1.54) \pm (0.12, 0.12) {\rm\ mas\ yr}^{-1}.$$

To compare this measurement to the geocentric relative proper motion, $\boldsymbol{\mu}_{\rm rel,geo}$, determined from the light curve model (Gould et al. 2006; Bennett et al. 2015), we must convert it to the inertial geocentric frame that moves at the velocity of the Earth at the light curve peak. This conversion will be explained in the next section as it is required in our Bayesian analysis.

### 4.2. Lens Mass from H-band Measurements

The measured magnitude in $H$ band is converted into absolute magnitude using

$$M_H = H_L - A_H - {\rm DM} = H_L - A_H - 5\log\frac{D_L}{10\ {\rm pc}} \quad (5)$$

where $A_H$ is the extinction along the line of sight and DM the distance modulus. At these Galactic coordinates, $(l, b) = (0.6769, -4.7402)°$, we assume a total extinction up to the Galactic Center of $A_H = 0.374 \pm 0.02$, as presented in Bennett et al. (2015), who use the method of Bennett et al. (2010) and the dereddened red clump magnitudes of Nataf et al. (2013). To calculate the extinction at a given distance $D_L$, we use the following expression:

$$A_{H,L} = \frac{1 - e^{-|D_L/(h_{\rm dust} \sin b)|}}{1 - e^{-|D_{\rm GC}/(h_{\rm dust} \sin b)|}} A_H \quad (6)$$

where $h_{\rm dust}$ is the dust scale height, $h_{\rm dust} = 0.10 \pm 0.02$ kpc (Bennett et al. 2015) and $D_{\rm GC}$ is the Galactic Center distance, assumed to be 8.01 kpc, since at these coordinates DM = 14.517 (Nataf et al. 2013).

We want to correlate the lens flux measurement (Equation (5)) with a calibrated population of main sequence stars. To do so, we adopt isochrones from An et al. (2007) and Girardi et al. (2002) that provide a mass–luminosity function for different ages and metallicities of main sequence stars. We choose ages from 500 Myrs to 10 Gyrs, and metallicities within the range $0.0 \leqslant {\rm [Fe/H]} \leqslant +0.2$. These isochrones, plotted as a function of the lens distance $D_L$ instead of its mass, assume that we know the Einstein radius $\theta_E$ of the microlensing system, which constrains the mass–distance relation.

For the previous analyses using adaptive optics (e.g., Batista et al. 2014; Bennett et al. 2014), we used the value of $\theta_E$ from the microlensing light curve modeling to build the isochrone profiles as a function of the distance of the lens star. However, in the case of OGLE-2005-BLG-169Lb, we also measure the heliocentric relative proper motion, $\boldsymbol{\mu}_{\rm rel,helio}$, between the source and the lens since the two objects are resolved more than eight years after the microlensing event. We then want to include this measurement in the determination of the lens distance, especially since the Einstein radius determined from the light curve (Gould et al. 2006; Bennett et al. 2015, see also Section 2), $\theta_E = \theta_* t_E/t_*$, has large uncertainties ($\sim 10$%) due to the fact that the data only covered the caustic exit of the event. Indeed, $\theta_E$ can also be constrained by the geocentric relative proper motion, $\theta_E = t_E \mu_{\rm rel,geo}$. To update this value, we perform a Bayesian analysis using both the constraints from the 2005 microlensing light curve ($t_E$, $\theta_E$, $f_S$) modeled by Bennett et al. (2015) and those from the Keck measurements ($\boldsymbol{\mu}_{\rm rel,helio}$, $f_L/f_S$) as prior distributions to determine the properties of the lensing system with the highest probabilities ($\theta_E$, $t_E$, $\boldsymbol{\mu}_{\rm rel,geo}$, $D_L$, $D_S$, $M_{\rm host}$, $m_{\rm planet}$, $r_\perp$). The constraints from the light curve correspond to the MCMC average parameters shown in Table 1. The prior distributions are based on the following $1\sigma$ ranges: $t_{\rm E} = 41.8 \pm 2.9$ days, $\theta_{\rm E} = 0.965 \pm 0.094$ mas, $H_{\rm S} = 18.81 \pm 0.08$, $\mu_{\rm rel,helio}$(R.A., decl.) $= (5.63, 4.87) \pm (0.12, 0.12)$ mas yr$^{-1}$, $f_{\rm L}/f_{\rm S} = 1.75 \pm 0.03$, and $D_{\rm S} = 8.5 \pm 1.5$ kpc.

The source is most likely located in the bulge, but its distance is not precisely known from measurements. We then consider a mean distance that takes into account the distribution of stars in the bulge and the maximization of the microlensing rate. This evaluation is affected by the increase of the solid angle along the line of sight, which increases the Einstein radius and pushes distances behind the galactic bar center. Additionally, the decrease in the brightness of stars compensates for a minor part of the latter effect by favoring closer sources. According to Nataf et al. (2013), at these coordinates the bar is located at $D = 8.01 \pm 1.06$ kpc, which would result in the range $8.25 \pm 1.1$ kpc for the source distance estimate. However, the distance of the Galactic Center has not reached a concensus yet and is often assumed to be farther than the one given by Nataf et al. (2013), 8.20 kpc (e.g., $8.27 \pm 0.29$ kpc in Schönrich 2012 and $8.4 \pm 0.4$ kpc in Ghez et al. 2008). Moreover, when using a galactic model based on Robin et al. (2003), the source is more likely to be situated at 8.7 kpc when following the same strategy of maximizing the microlensing rate. The source distribution used in the companion *HST* paper (Bennett et al. 2015) is centered on this value. In this work, we chose a source distance of $D_{\rm S} = 8.5 \pm 1.5$ kpc. Thus we are conservative by covering a large range of possibilities. The impact of the source distance variation (between 8.25, 8.5, and 8.7 kpc) on the lens distance is less than 5%, i.e., within $1\sigma$ error bar on the resulting lens mass. Within the next two years, *Gaia* will provide accurate description of the galactic disk and bulge. Although this particular source will not be in the *Gaia* catalog because it is too faint, we can expect a significant improvement in estimating the microlensing source distances.

Two-hundred thousand combinations of $(t_{\rm E,in}, \theta_{\rm E,in}, D_{\rm S}, \mu_{\rm rel,helio})$ were tested, providing resulting probabilities for $(\theta_{\rm E,out}, t_{\rm E,out}, \mu_{\rm rel,geo}, D_{\rm L}, M_{\rm host}, m_{\rm planet}, r_\perp)$. To convert the heliocentric relative proper motion into a geocentric frame, we need to assume a starting value for the distance of the lens, $D_{\rm L,0}$, since the offset between them is a function of the relative distance between the lens and the source:

$$\mu_{\rm rel,geo} = \mu_{\rm rel,helio} - \Delta\mu$$

where

$$\Delta\mu = \frac{\pi_{\rm rel} V_{\oplus,\perp}}{\rm AU} = \left(\frac{1}{D_{\rm L}} - \frac{1}{D_{\rm S}}\right) V_{\oplus,\perp}$$

(Dong et al. 2009).

Here $V_{\oplus,\perp}$ is the velocity of the Earth projected on the sky at the time of the microlensing event (i.e., at $t_0$), in north and east coordinates (Gould 2004):

$$V_{\oplus,\perp} = (3.1, 18.5) \text{ km s}^{-1} = (0.65, 3.9) \text{ AU yr}^{-1}$$

As the initial $D_{\rm L,0}$ value that we use to determine $\mu_{\rm rel,geo}$ is a guess, we need to perform several loops to converge to a stable and appropriate lens distance. For each tested value of $\mu_{\rm rel,helio}$, we obtain a value of $\mu_{\rm rel,geo}$ that gives $\theta_{\rm E} = \mu_{\rm rel,geo} t_{\rm E}$. This $\theta_{\rm E}$ is then used in the mass–distance relation that enables the mass–luminosity relations (isochrones) to be crossed with the magnitude profile of the lens $M_H$ as a function of the lens distance $D_{\rm L}$. Our code calculates the intersection $D_{\rm L}$ between these two curves and re-injects this distance as a starting lens distance $D_{\rm L,0}$ if it differs from the previous one by more than 0.2%. This recursive method converges to a stable value of $D_{\rm L}$ that is consistent with the parameters from the microlensing light curve, the Keck measurement of the relative proper motion, as well as the measured magnitude of the lens.

The posterior distributions of $(D_{\rm L}, M_{\rm host}, m_{\rm planet}, r_\perp)$ from the Bayesian analysis are presented in Figure 4. As a comparison, the 90% confidence ranges given by Gould et al. (2006) are shown by shaded blue areas. These two solutions for the lens system properties are consistent, with much smaller error bars for the updated values thanks to the additional Keck constraints. The error bars from the Bayesian analysis have been combined with the uncertainties contained in Equation (5) (extinction and lens magnitude) and the dispersion of isochrones from different populations, as shown in Figure 5. The intersection between the isochrones and the lens magnitude occurs at an absolute magnitude of $M_H = 5.0^{+0.6}_{-0.4}$ and a distance $D_{\rm L} = 4.0 \pm 0.4$ kpc, when using the set of microlensing parameters with the highest probability provided by the Bayesian analysis. This corresponds to a geocentric relative proper motion of $\mu_{\rm geo} = 7.0 \pm 0.2$ mas yr$^{-1}$. This measurement is consistent with the $\mu_{\rm rel,geo}$ from the light curve, although the Keck values are in the extreme low part of this previous $\mu_{\rm rel,geo}$ distribution. As explained in Section 2, these new values of $\mu_{\rm rel,geo}$ exclude a range of mass ratios and favor the light curve model that gives $q = 6.15 \pm 0.40 \times 10^{-5}$. The final ranges for the planetary system properties are given in Table 2, combining all sources of uncertainties.[5]

### 4.3. Comparison to HST Wide Field Camera 3-Ultraviolet-Visible (WFC3-UV) Measurements

OGLE-2005-BLG-169 was observed on 2011 October 19 (i.e., 6.4678 years after the microlensing event) with the *HST* using the WFC3-UV instrument. An independent analysis is presented in a companion paper, Bennett et al. (2015). Their point-spread function fitting procedure unambiguously detects two stars at the position of the target (the lens and the source), which allows them to calculate the two-dimensional relative proper motion between the lens and the source. Their measurement is consistent with our Keck measurement and both $l$ and $b$ components are within $1\sigma$ of our values:

$$\mu_{\rm rel,helio}(l, b)[HST] = (7.39 \pm 0.20, 1.33 \pm 0.23) \text{ mas yr}^{-1}.$$

If we combine the measurements of the lens–source relative projected separation at two different epochs using Keck and *HST*, we can estimate the closest projected distance between these stars at the time of the microlensing event (at $t_0$). We obtain

$$(\Delta l, \Delta b) = (2 \pm 6, -7 \pm 7) \text{ mas},$$

---

[5] If we consider a random orbit, we can assume a semimajor axis $a \sim 1.15 \times r_\perp \sim 4.0$ AU. The 1.15 factor comes from the assumption that, for circular orbits, the distribution of $\cos(\beta)$ is uniform, where $r_\perp = a* \sin(\beta)$ and $a$ is the semimajor axis. Thus, the median of $\cos(\beta)$ is 0.5. Therefore, $\langle a/r_\perp\rangle_{\rm med} = 1/\sqrt{(1 - 0.5^2)} \sim 1.15$.

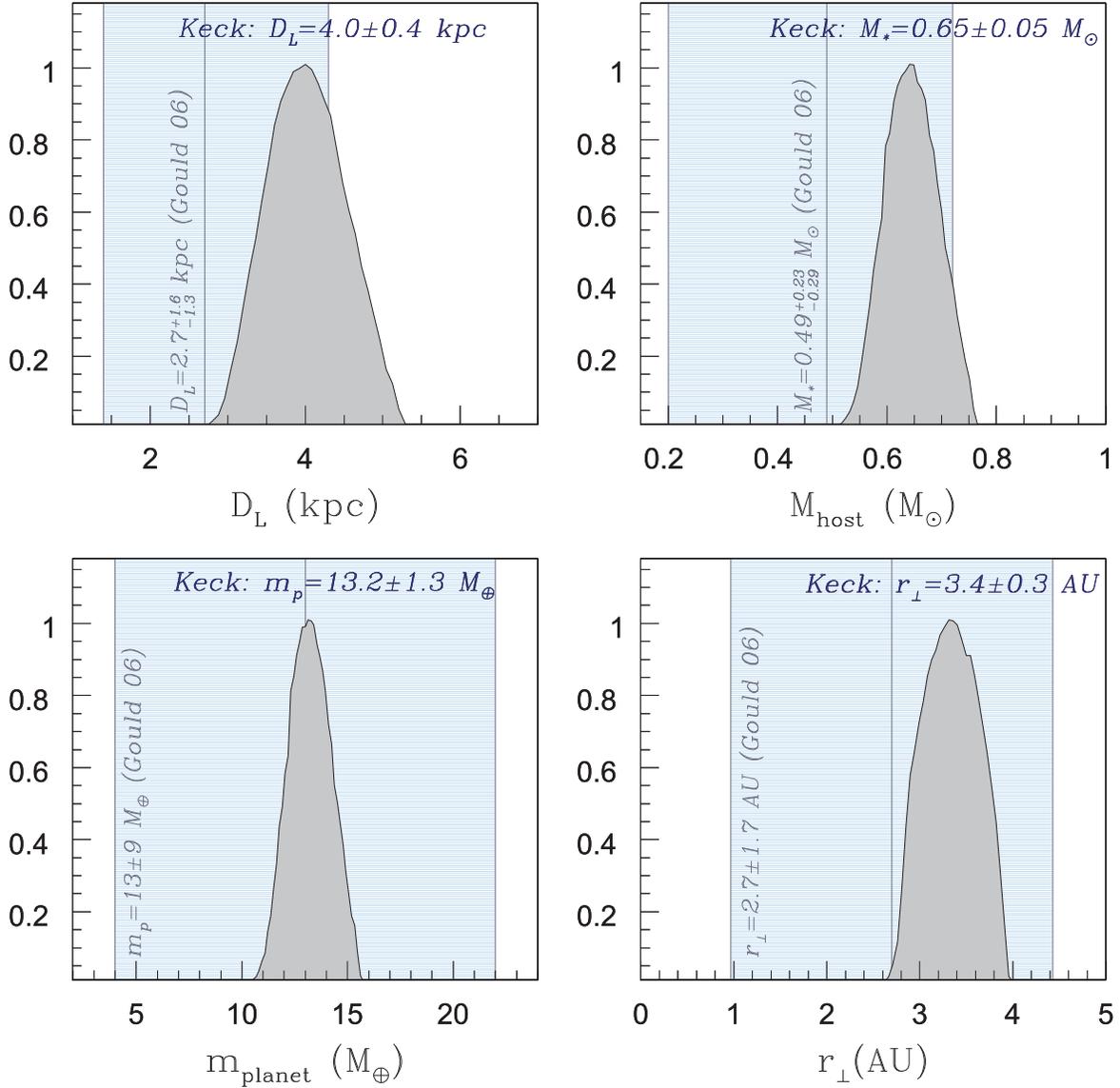

**Figure 4.** Final distributions for the lens system parameters ($D_L$, $M_{host}$, $m_{planet}$, $r_\perp$) from the Bayesian analysis using the Keck measurements as priors in gray. Blue: 90% confidence ranges from Gould et al. (2006).

which is consistent within $1\sigma$ with the estimated lens–source separation from the modeling. Our Bayesian analysis provides a posterior value for the Einstein radius of $\theta_{E,out} = 0.838$ mas, and thus, according to the minimum separation $u_0 = 0.001267$ from Bennett et al. (2015) at $t_0$, $\theta_{E,out} u_0 = 0.00106$ mas.

The $3\sigma$ range of the estimated separation of the stars at $t_0$ ensures that the detected bright object in the Keck images was less than 30 $\theta_E$ away from the source. One wonders whether this star is indeed the lens or a companion to the lens. However, if the detected star were not the lens, it would obviously be much bigger than the lens (to be seen while the lens is not seen) and would have created distortions in the light curve. Indeed, since it is a high magnification event, one can expect a high sensitivity to lens companions, and in the present case such a massive object within 30 $\theta_E$ would have been detectable in the light curve. It would induce a shear, $\gamma = q/s^2 > 10^{-3}$, on the lens gravitational field that would generate a caustic at the center of magnification of the lens system (Chang & Refsdal 1979, 1984).

### 4.4. Confirmation of the Gould et al. (2006) Parallax Estimate

Although the parallax signal detected in the discovery paper was too weak to be considered as a strong constraint for the determination of the lens mass, we can compare their estimated microlens parallax to the one induced by Keck measurements. Indeed they derived $\pi_{E,\parallel} = -0.086 \pm 0.261$ from the light curve, where $\pi_{E,\parallel} = \pi_E \cos\psi$, and $\psi$ is the angle between the direction of lens–source relative motion and the position of the Sun at $t_0$ projected on the plane of the sky (Gould 2004).

The parallel component of the microlens parallax can be expressed as

$$\pi_{E,\parallel} = \boldsymbol{u}_{Sun} \cdot \boldsymbol{\mu}_{rel,geo} \frac{\pi_{rel} t_E}{\theta_E^2}$$

where $\boldsymbol{u}_{Sun}$ is the unity vector of the Sun–Earth projected separation at $t_0$, $\boldsymbol{u}_{Sun}(E, N) = (-0.995, +0.097)$ AU, $\boldsymbol{\mu}_{rel,geo} = (5.16, 4.79)$ mas yr$^{-1}$ from the Keck measurements, and $\pi_{rel} = 0.11$ mas AU$^{-1}$.

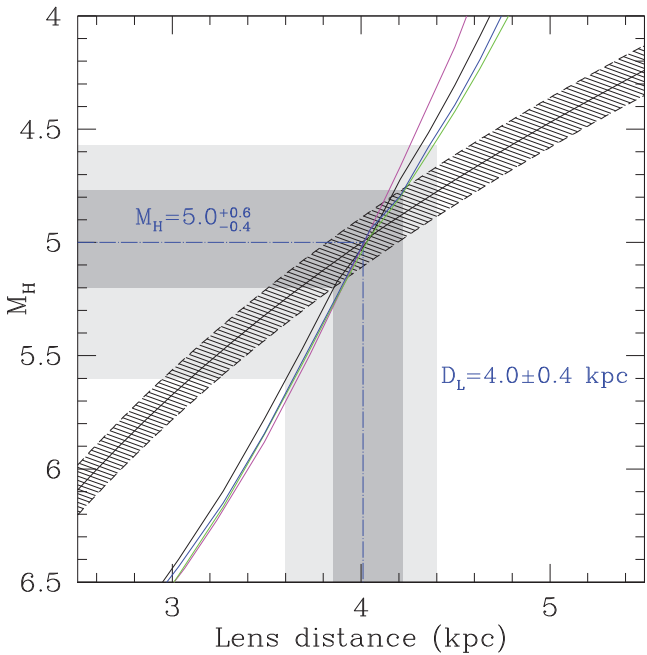

**Figure 5.** Estimate of the lens absolute magnitude measured by Keck AO observations in $H$ band (black curves, dashed area including error bars on the magnitude and extinction). The colored curves are isochrones for main sequence stars from An et al. (2007) and Girardi et al. (2002), with the following ages and metallicities: 10 Gyr, [Fe/H] = 0.0 (pink), 4 Gyr, [Fe/H] = 0.0 (black), 0.5 Gyr, [Fe/H] = 0.1 (light green), and 4 Gyr, [Fe/H] = 0.2 (blue). The intersection between the Keck measurements and the 10 Gyr isochrone defines the solution, while the other isochrone populations represent the mass–luminosity function dispersion. The dark gray areas take into account both this dispersion and the uncertainties on the absolute magnitude of the lens. The light gray range includes the additional uncertainties coming from the Bayesian analysis.

**Table 2**
Physical Parameters with $1\sigma$ Ranges

| Parameter | Units | Value |
|---|---|---|
| $D_L$ | kpc | $4.0 \pm 0.4$ |
| $M_{host}$ | $M_\odot$ | $0.65 \pm 0.05$ |
| $m_{planet}$ | $M_\oplus$ | $13.2 \pm 1.3$ |
| $r_\perp$ | AU | $3.4 \pm 0.3$ |
| $a$ | AU | $\sim 4.0$ |

Combined with our Bayesian posteriors, $t_{E,out} = 43.63$ days and $\theta_{E,out} = 0.838$ mas, we deduce a microlens parallax of $\pi_{E,\|} = -0.090$ which is very close to the Gould et al. (2006) estimate.

## 5. CONCLUSION

During a microlensing event, a distant source star is temporarily magnified by the gravitational potential of a foreground object. If this lensing object is a planetary system, additional perturbations can be created by the planetary companions in addition to the main features due to the host star. The inverse problem of finding the properties of the lensing system (planet/star mass ratio, star–planet projected separation) from an observed light curve is a complex nonlinear one within a wide parameter space. Moreover, when the microlens parallax cannot be measured or when its signal is too weak or degenerate with other parameters, galactic models are needed to extract the physical parameters of the planetary system in the last stage of the analysis. Nevertheless, it is possible to confirm the models and refine the properties of the planetary systems detected by microlensing. An independent method to constrain the lens star mass consists in measuring its flux directly (or an upper limit) using high angular resolution imaging to separate the source-lens from the subarsec blended stars. Even stronger constraints are obtained when the source and the lens stars are separated enough so that their flux ratio and relative proper motion could be measured. Hence, this new measurement of $\mu_{rel,helio}$, as well as the measured brightness of the lens in $H$ band, enabled the mass and distance of the system to be updated: a Uranus planet orbiting a K5-type main sequence star: $m_p = 13.2 \pm 1.3 M_\oplus$, $M_* = 0.65 \pm 0.05 M_\odot$, separated by $a_\perp = 3.4 \pm 0.3$ AU, at the distance $D_L = 4.0 \pm 0.3$ kpc from us.

It is the first time that a planetary signal in a microlensing light curve is confirmed by additional observations taken after the event. Moreover, the measurement of the lens–source relative proper motion yields an estimate of the parallax effects that would have affected the light curve in 2005. This estimate is in agreement with the one given by Gould et al. (2006), whereas the parallax signature in the light curve was weak due to sparse data.

Such measurements pioneered with Keck and *HST* (see also Bennett et al. 2015) will become routine for the microlensing surveys on board *Euclid* and *WFIRST* (Bennett & Rhie 2002; Penny et al. 2013; Spergel et al. 2013).

This work was supported by a NASA Keck PI Data Award, administered by the NASA Exoplanet Science Institute. Data presented herein were obtained at the W. M. Keck Observatory from telescope time allocated to the National Aeronautics and Space Administration through the agency's scientific partnership with the California Institute of Technology and the University of California. The Observatory was made possible by the generous financial support of the W. M. Keck Foundation. V. B. was supported by the Programme National de Planetologie and the DIM ACAV, Region Ile de France. U. B. and J. P. B. were supported by PERSU Sorbonne Université. D. P. B. and A. B. were supported by NASA through grants from the Space Telescope Science Institute (# 12541 and 13417), as well as grants NASA-NNX12AF54G and NSF AST-1211875. A. G. was supported by NSF grant AST 1103471. The OGLE project has received funding from the European Research Council under the European Community's Seventh Framework Programme (FP7/2007-2013)/ERC grant agreement No. 246678 to A. U.

## REFERENCES

An, D., Terndrup, D. M., Pinsonneault, M. H., et al. 2007, ApJ, 655, 233
Batista, V., Beaulieu, J.-P., Gould, A., et al. 2014, ApJ, 780, 54
Batista, V., Gould, A., Diesters, S., Dong, S., et al. 2011, A&A, 529, 102
Beaulieu, J.-P., Bennett, D. P., Fouqué, P., et al. 2006, Natur, 439, 437
Bennett, D. P., Alcock, C., Allsman, R., et al. 1993, BAAS, 25, 1402
Bennett, D. P., Anderson, J., & Gaudi, B. S. 2007, ApJ, 660, 781
Bennett, D. P., Batista, V., Bond, I. A., et al. 2014, ApJ, 785, 155
Bennett, D. P., Bhattacharya, A., Anderson, J., et al. 2015, ApJ, 808, 169
Bennett, D. P., & Rhie, S. H. 1996, ApJ, 472, 660
Bennett, D. P., & Rhie, S. H. 2002, ApJ, 574, 985
Bennett, D. P., Rhie, S. H., Nikolaev, S., et al. 2010, ApJ, 713, 837
Boyajian, T. S., van Belle, G., & von Braun, K. 2014, AJ, 147, 47
Chang, K., & Refsdal, S. 1979, Natur, 282, 561
Chang, K., & Refsdal, S. 1984, A&A, 130, 157

Diolaiti, E., Bendinelli, O., Bonaccini, D., et al. 2000, A&AS, 147, 335
Dong, S., Gould, A., Udalski, A., et al. 2009, ApJ, 695, 970
Gaudi, B. S., Bennett, D. P., Udalski, A., et al. 2008, Sci, 319, 927
Ghez, A. M., Salim, S., Weinberg, N. N. A., et al. 2008, ApJ, 689, 1044
Girardi, L., Bertelli, G., Bressan, A., et al. 2002, A&A, 391, 195
Gould, A. 2004, ApJ, 606, 319
Gould, A. 2014, JKAS, 47, 279
Gould, A., & Loeb, A. 1992, ApJ, 396, 104
Gould, A., Udalski, A., An, D., et al. 2006, ApJL, 644, L37
Griest, K., & Safizadeh, N. 1998, ApJ, 500, 37
Han, C., Udalski, A., Hikasa, K., et al. 2013, ApJL, 762, L28
Ida, S., & Lin, D. N. C. 2004, ApJ, 616, 567
Janczak, J., Fukui, A., Dong, S., et al. 2010, ApJ, 711, 731
Kervella, P., Thévenin, F., di Folco, E., & Ségransan, D. 2004, A&A, 426, 297
Kubas, D., Beaulieu, J.-P., Bennett, D. P., et al. 2012, A&A, 540, A78
Lissauer, J. J. 1993, ARA&A, 31, 129
Mao, S., & Paczyński, B. 1991, ApJL, 374, L37
Minniti, D., Lucas, P. W., Emerson, J. P., et al. 2010, NewA, 15, 433
Muraki, Y., Han, C., Bennett, D. P., et al. 2011, ApJ, 741, 22
Nataf, D. M., Gould, A., Fouqué, P., et al. 2013, ApJ, 769, 88
Penny, M. T., Kerins, E., Rattenbury, N., et al. 2013, MNRAS, 434, 2
Robin, A. C., Reylé, C., Derrière, S., & Picaud, S. 2003, A&A, 409, 523
Schönrich, R. 2012, MNRAS, 427, 274
Spergel, D., Gehrels, N., Breckinridge, J., et al. 2013, arXiv:1305.5422
Yoo, J., DePoy, D. L., Gal-Yam, A., et al. 2004, ApJ, 603, 139